\documentclass{PoS}

\title{No-Go Theorem of Leibniz Rule and Supersymmetry on the Lattice}

\ShortTitle{No-Go Theorem of Leibniz Rule and Supersymmetry on the Lattice}

\author{Mitsuhiro Kato\\
        Institute of Physics, University of Tokyo, Komaba, 
        Meguro-ku, Tokyo 153-8902, Japan\\
        E-mail: \email{kato@hep1.c.u-tokyo.ac.jp}
        }
\author{Makoto Sakamoto\footnote{speaker}\ \footnote{
        The authors would like to thank to K. Nagata and N. Ukita for
        useful discussions. 
        This work is supported in part by the Grant-in-Aid for Scientific 
        Research (No.19540272 (M.K.), No.18540275 (M.S.) and No.17540242 (H.S.))
        by the Japanese Ministry of Education, Science, Sports and Culture. 
        }\\
        Department of Physics, Kobe University, Nada-ku, 
        Hyogo  657-8501, Japan\\
        E-mail: \email{dragon@kobe-u.ac.jp}
        }
\author{Hiroto So\\
        Department of Physics,  Ehime University, Bunkyou-chou 2-5, 
        Matsuyama 790-8577, Japan\\
        E-mail: \email{so@phys.sci.ehime-u.ac.jp}
        }
\abstract{
An obstacle to realize supersymmetry on a lattice is the 
breakdown of Leibniz rule.
We give a proof of a no-go theorem 
that it is impossible to construct a lattice field theory 
in an infinite lattice volume with any nontrivial field products
and difference operators that satisfy the
following three properties:
(i) translation invariance, (ii) locality and (iii) Leibniz rule.
We then propose a way to escape from the no-go theorem
by introducing infinite flavors, and present a lattice
model of $N=2$ supersymmetric quantum mechanics equipped
with the full exact supersymmetry.
}
\FullConference{The XXVI International Symposium on Lattice Field Theory\\
		 July 14-19 2008\\
		 Williamsburg, Virginia, USA}
\begin{document}
%%%%%%%%%%%%%%%%%%%%%%%%%%%%%%%%%%%%%%%%%%%%%%%%%%%%%%%%%%%%%%%%
%%%%%%%%%%%%%%%%%%%%%%%%%%%%%%%%%%%%%%%%%%%%%%%%%%%%%%%%%%%%%%%%
%
%
%
%
%
%
%%%%%%%%%%%%%%%%%%%%%%%%%%%%%%%%%%%%%%%%%%%%%%%%%%%%%%%%%%%%%%%%
%%%%%%%%%%%%%%%%%%%%%%%%  Section 1  %%%%%%%%%%%%%%%%%%%%%%%%%%%
%%%%%%%%%%%%%%%%%%%%%%%%%%%%%%%%%%%%%%%%%%%%%%%%%%%%%%%%%%%%%%%%
\section{Introduction}
%%%%%%%%%%%%%%%%%%%%%%%%%%%%%%%%%%%%%%%%%%%%%%%%%%%%%%%%%%%%%%%%
%
%
%
Over the last thirty years, a considerable number of attempts
have been made to construct lattice supersymmetric 
models \cite{Feo03,Kaplan04,Giedt06}.
However, none of them have been fully successful.
To see an obstacle in realizing supersymmetry on a lattice,
let us consider supersymmetry invariance in the continuum,
which may be  best seen in the superfield formulation.
We first note that an infinitesimal supersymmetry transformation 
for any superfield $X(x,\theta)$ is essentially written in the form
$\delta_{\epsilon} X(x,\theta) = \epsilon(\partial_{\theta} + \theta
\partial_{x})\, X(x,\theta)$, 
where $\partial_{x}$ and $\partial_{\theta}$ denote the
derivatives with respect to the spacetime coordinate $x$ and 
the grassmann coordinate $\theta$, respectively, and
$\epsilon$ is an infinitesimal grassmann parameter.
%where $x$ and $\theta$ denote
%spacetime and grassmann coordinates, respectively, and
%$\epsilon$ is an infinitesimal grassmann parameter.
Then, the action invariance under the transformation
can easily be shown as
%The essential feature of the action invariance under infinitesimal
%supersymmetry transformations will be explained as follows:

\vspace{-5mm}
%
%%%%%%%%%%%%%%%%
\begin{eqnarray}
\label{action_invariance}
S(\Phi + \delta_{\epsilon}\Phi) - S(\Phi)
  &\equiv& \int\!\! dx\int\!\! d\theta\ \big\{
        {\cal{L}}(\Phi(x,\theta)+\delta_{\epsilon}\Phi(x,\theta))
        -{\cal{L}}(\Phi(x,\theta)) \big\} \nonumber\\
  &=& \int\!\! dx\int\!\! d\theta\ \delta_{\epsilon}\!{\cal{L}}(x,\theta) 
        \nonumber\\
%  &=& \int\!\! dx\int\!\! d\theta\ \epsilon\big[ \partial_{\theta}
%        + \theta \partial_x\big] {\cal{L}}(x,\theta) 
%        \nonumber\\ 
  &=& 0\, ,      
\end{eqnarray}
%%%%%%%%%%%%%%%%
%

\vspace{-2mm}
\noindent
where $\Phi$ stands for a superfield. 
The last equality immediately follows from the fact that
$\int\!\! dx\ \partial_{x} = 0$ and
$\int\!\! d\theta\ \partial_{\theta} = 0$.  
%$x$ and $\theta$ stand for
%spacetime and grassmann coordinates, respectively, and 
%$\epsilon$ is an infinitesimal grassmann parameter.
%The third line in eq.(\ref{action_invariance})
%shows that infinitesimal supersymmetry transformations 
%of the action are given by total derivatives of spacetime and grassmann
%coordinates.
%This fact immediately leads to the action invariance
%by use of the properties:
%$\int\!\! dx\ \partial_{x} = 0$ and
%$\int\!\! d\theta\ \partial_{\theta} = 0$.  
%
%%%%%%%%%%%%%%%%
%\begin{eqnarray}
%%
%\label{total_derivative}
%\int\!\! dx\ \partial_{x} = 0\,, \qquad
%\int\!\! d\theta\ \partial_{\theta} = 0\,.   
%%
%\end{eqnarray}
%%%%%%%%%%%%%%%%
%

In lattice formulation, $x$, $\partial_{x}$ and $\int \!\!dx$
should be replaced by $n$, $\nabla$ and $\sum_{n}$, 
respectively, where $n$ and $\nabla$ denote a position and
a difference operator on a lattice.
Since ordinary difference operators satisfy the relation
$\sum_{n} \nabla = 0$, most of the properties in the
continuum turn out to hold even on a lattice.
A nontrivial step, however, exists.
To go from the first line to the second in eq.(\ref{action_invariance}),
we need a Leibniz rule for the supersymmetry transformations,
i.e. $\delta_{\epsilon}(\Phi\Psi) = (\delta_{\epsilon}\Phi)\Psi 
+ \Phi(\delta_{\epsilon}\Psi)$.
The Leibniz rule is trivially satisfied in the continuum 
but not on a lattice.
This is because the supersymmetry transformations include
spacetime derivatives, which should be replaced by some
difference operators on a lattice, and naive difference
operators are known to violate the Leibniz 
rule \cite{SLAC76, Dondi-Nicolai77, Bouguenaya-Fairlie86, So-Ukita99}.
Thus, it will be important to clarify the question whether
the Leibniz rule can be realized on a lattice.

Our answer is negative.
In the next section, we give a proof of the no-go 
theorem \cite{Taming}:

%%
%%
%%%%%%%%%%%%%%%%%%%%%%%%  No-Go Theorem  %%%%%%%%%%%%%%%%%%%%%%%
%\begin{theorem}
%\label{theorem}
%It is impossible to construct a lattice field theory in an infinite
%lattice volume with a nontrivial product rule (\ref{product})
%and a difference operator (\ref{diff}) that satisfy the
%following three properties:
%(i) translation invariance, (ii) locality and (iii) Leibniz rule.
%%
%\end{theorem}
%%%%%%%%%%%%%%%%%%%%%%%%%%%%%%%%%%%%%%%%%%%%%%%%%%%%%%%%%%%%%%%%
%%
%

\vspace{1mm}
%
%
%%%%%%%%%%%%%%%%%%%%%%%  No-Go Theorem  %%%%%%%%%%%%%%%%%%%%%%%
\noindent
{\bf No-Go Theorem.}\ \ 
{\it 
It is impossible to construct a lattice field theory in an infinite
lattice volume with any nontrivial field products
and difference operators that satisfy the
following three properties:
(i) translation invariance, (ii) locality and (iii) Leibniz rule.}
%%%%%%%%%%%%%%%%%%%%%%%%%%%%%%%%%%%%%%%%%%%%%%%%%%%%%%%%%%%%%%%
%
%
\vspace{1mm}

\noindent
We, however, propose a way to escape from the no-go theorem 
and present a lattice formulation equipped with a Leibniz rule
in Section 3.
Then, we explicitly construct a lattice model of $N=2$
superesymmetric quantum mechanics and discuss the full exact 
supersymmetric properties from various points of view
in Section 4.
Section 5 is devoted to summary and discussions.
%Most of the results in Sections 2 and 3 have been given
%in Ref.\cite{Taming} but the results in Section 4 are new.

%
%
%
%%%%%%%%%%%%%%%%%%%%%%%%%%%%%%%%%%%%%%%%%%%%%%%%%%%%%%%%%%%%%%%%
%%%%%%%%%%%%%%%%%%%%%%%%  Section 2  %%%%%%%%%%%%%%%%%%%%%%%%%%%
%%%%%%%%%%%%%%%%%%%%%%%%%%%%%%%%%%%%%%%%%%%%%%%%%%%%%%%%%%%%%%%%
\section{A Proof of the No-Go Theorem}
%%%%%%%%%%%%%%%%%%%%%%%%%%%%%%%%%%%%%%%%%%%%%%%%%%%%%%%%%%%%%%%%
%
%
%
In this section, we present a proof of the no-go theorem 
stated in the Introduction.
In order to analyze a Leibniz rule on a lattice, we must
generalize a product between fields and a difference
operator on a field.
A lattice field product between $\phi_{n}$ and $\psi_{n}$
and a difference operator on a field $\phi_{n}$ are defined,
respectively, as

\vspace{-3mm}
%
%%%%%%%%%%%%%%%%
\begin{equation}
\label{product&diff}
(\phi \cdot \psi )_{n} \equiv
   \sum_{l,m} C_{lmn}\,\phi_{l}\psi_{m}\,,\quad
(D\phi)_{n} \equiv \sum_{m} D_{mn}\,\phi_{m}\,,
\end{equation}
%%%%%%%%%%%%%%%%
%

\vspace{-1mm}
\noindent
where the indices $l, m, n$ denote positions on a lattice
of an infinite size.
%The normal product of lattice fields at the same point
%corresponds to the choice 
%$C_{lmn} = \delta_{l,n}\,\delta_{m,n}$
%as the field product.
%Here, we allow the lattice fields $\phi_{l}$ and $\psi_{m}$
%at different positions to be coupled in the field product 
%(\ref{product}).

%A difference operator on a field $\phi_{n}$ is defined as
%%
%%%%%%%%%%%%%%%%%
%\begin{equation}
%%
%\label{diff}
%(D\phi)_{n} \equiv \sum^{\infty}_{m=-\infty}
%    D_{mn}\,\phi_{m}\,.
%%
%\end{equation}
%%%%%%%%%%%%%%%%%
%
%The difference operator for a constant field is required
%to vanish, i.e.
%%
%%%%%%%%%%%%%%%%%
%\begin{equation}
%%
%\label{condition_diff}
%\sum^{\infty}_{m=-\infty}D_{mn} = 0\,.
%%
%\end{equation}
%%%%%%%%%%%%%%%%%
%
%Three familiar examples for the forward, backward and
%symmetric difference operators are given by
%$D^{(+)}_{mn}=\delta_{m,n+1} - \delta_{m,n}$,
%$D^{(-)}_{mn}=\delta_{m,n} - \delta_{m,n-1}$ and
%$D^{(+)}_{mn}=1/2(\delta_{m,n+1} - \delta_{m,n-1})$, 
%respectively.

The coefficients $C_{lmn}$ and $D_{mn}$ cannot be arbitrary.
The difference operator for a constant field should
vanish, i.e. $\sum_{m}D_{mn} = 0$.
%
%%%%%%%%%%%%%%%%%
%\begin{equation}
%%
%\label{condition_diff}
%\sum^{\infty}_{m=-\infty}D_{mn} = 0\,.
%%
%\end{equation}
%%%%%%%%%%%%%%%%%
%
We require that they are translationally invariant, 
such that
$C_{lmn} = C(l-n,m-n)$ and $D_{mn} = D(m-n)$.
%%
%%%%%%%%%%%%%%%%%
%\begin{equation}
%%
%\label{translation_invariance}
%C_{lmn} = C(l-n,m-n),\quad D_{mn} = D(m-n)\,.
%%
%\end{equation}
%%%%%%%%%%%%%%%%%
%
We further require the field product and the difference
operator to be local.
The locality property is important in constructing
local field theories after the continuum limit.
To make the locality manifest, we define Fourier
transform of the coefficients $C(k,l)$ and $D(m)$ by

\vspace{-6mm}
%
%%%%%%%%%%%%%%%%
\begin{eqnarray}
\label{Fourier_transform}
\hat{C}(v,w) 
   \equiv \sum_{k,l}
              C(k,l)\,v^{k}\,w^{l}\,,\quad
%\label{Fourier_D}
\hat{D}(z) 
   \equiv \sum_{m}
              D(m)\,z^{m}\,,             
\end{eqnarray}
%%%%%%%%%%%%%%%%
%

\vspace{-2mm}
\noindent
where $v, w, z$ are $S^{1}$-variables given by
$v=e^{ip}, w=e^{iq}, z=e^{ir}\ (0\le p, q, r < 2\pi)$.
We note that the condition $\sum_{m}D_{mn} = 0$ can be replaced by
an \lq\lq initial\rq\rq\ condition

\vspace{-3mm}
%
%%%%%%%%%%%%%%%%
\begin{equation}
\label{condition_diff2}
\hat{D}(1) = 0\,.
\end{equation}
%%%%%%%%%%%%%%%%
%

\vspace{-1mm}
\noindent
A crucial observation to prove the no-go theorem is that 
the locality of the field product and the difference operator
allows $v, w, z$ to extend to an annulus domain 
$1-\epsilon < |v|, |w|, |z| < 1+\epsilon$ with a positive
constant $\epsilon$ and then $\hat{C}(v,w)$ and $\hat{D}(z)$
can be regarded as holomorphic functions on the annulus
domain \cite{Taming}.

We are now ready to present a proof of the no-go theorem.
The Leibniz rule $D(\phi\cdot\psi) = (D\phi)\cdot\psi
+\phi\cdot(D\psi)$ can be translated into a relation between
the field product and the difference operator as

\vspace{-6mm}
%
%%%%%%%%%%%%%%%%
\begin{equation}
\sum_{k} C_{lmk} D_{kn}  
  = \sum_{k} C_{kmn} D_{lk}  
    + \sum_{k} C_{lkn} D_{mk}\,. 
\label{leibniz}
\end{equation}
%%%%%%%%%%%%%%%%
%
%Multiplying the both sides of eq.(\ref{leibniz}) by
%$\sum_{l,m} v^{l-n}w^{m-n}$ and using the relations
%(\ref{translation_invariance}), we can rewrite the
%relation (\ref{leibniz}), in terms of the holomorphic
%functions $\hat{C}(v,w)$ and $\hat{D}(z)$, as
In terms of the holomorphic functions $\hat{C}(v,w)$ and 
$\hat{D}(z)$, the above relation can be rewritten into
a simple form

\vspace{-6mm}
%
%%%%%%%%%%%%%%%%
\begin{equation}
\hat{C}(v,w)\Big(\hat{D}(vw) - \hat{D}(v)- \hat{D}(w)\Big) = 0\, . 
\label{leibniz2}
\end{equation}
%%%%%%%%%%%%%%%%
%

\vspace{-2mm}
\noindent
It follows that
%
%%%%%%%%%%%%%%%%
\begin{equation}
\hat{D}(vw) - \hat{D}(v)- \hat{D}(w) = 0
\label{leibniz3}
\end{equation}
%%%%%%%%%%%%%%%%
%
on a restricted domain with $\hat{C}(v,w)\ne 0$.
We can, however, show that the relation (\ref{leibniz3})
holds on the annulus domain {\it without} the constraint
$\hat{C}(v,w)\ne 0$ by virtue of the holomorphy.
To see this, we first note that since $\hat{C}(v,w)$
is holomorphic, $\hat{C}(v,w)$ can vanish at most
at isolated points, otherwise $\hat{C}(v,w)$ should
vanish identically due to the identity theorem on
holomorphic functions.
It turns out that the equation (\ref{leibniz3}) holds
on the whole annulus domain
without the condition $\hat{C}(v,w)\ne 0$, again due
to the identity theorem on holomorphic functions.

A general solution to eq.(\ref{leibniz3}) is easily
found to be of the form

\vspace{-3mm}
%
%%%%%%%%%%%%%%%%
\begin{equation}
 \hat{D}(w) =  \beta \log w\,.
\label{log-sol}
\end{equation}
%%%%%%%%%%%%%%%%
%
The coefficient $\beta$, however, has to vanish in order
for $\hat{D}(w)$ to be holomorphic because the 
logarithmic function $\log w$ is not holomorphic on the 
annulus domain and hence is not local.
Thus, we conclude that there is no nontrivial difference operator
with  the properties of (i) translation invariance, (ii)
locality and (iii) Leibniz rule.
This completes the proof.

We would like to make a few comments on the locality.
Our proof implies that 
if we allow the difference operator to be non-local
(but still the field product to be local), the difference
operator (\ref{log-sol}) is a unique solution, which
is known as a SLAC-type derivative \cite{SLAC76}.
On the other hand, if we allow the field product to be
non-local, we can obtain local difference operators which
are solutions to eq.(\ref{leibniz2}).
For example, if we take $C_{lmn} = c$ ($=$constant),
which is highly non-local, then we find
$\hat{C}(v,w) = (2\pi i)^2 c\,\delta(v-1)\,\delta(w-1)$.
It then follows that any (local) difference operators
with the property (\ref{condition_diff2}) satisfy the
equation (\ref{leibniz2}) trivially.
Another example will be found in Ref.\cite{Dondi-Nicolai77}

%
%
%
%%%%%%%%%%%%%%%%%%%%%%%%%%%%%%%%%%%%%%%%%%%%%%%%%%%%%%%%%%%%%%%%
%%%%%%%%%%%%%%%%%%%%%%%%  Section 3  %%%%%%%%%%%%%%%%%%%%%%%%%%%
%%%%%%%%%%%%%%%%%%%%%%%%%%%%%%%%%%%%%%%%%%%%%%%%%%%%%%%%%%%%%%%%
\section{Escape from the No-Go Theorem}
%%%%%%%%%%%%%%%%%%%%%%%%%%%%%%%%%%%%%%%%%%%%%%%%%%%%%%%%%%%%%%%%
%
%
%
Although we have proved a no-go theorem about the Leibniz rule
on a lattice, it is still worthwhile continuing to search for
local difference operators and field products satisfying 
the Leibniz rule because they will 
make it possible to construct lattice models with 
the full exact supersymmetry.

A way to escape from the no-go theorem may be to introduce
multi-flavors in the system.\footnote{
Another approach to restore a Leibniz rule with a mild
non-commutativity has been proposed in Ref.\cite{Hokudai}.
}
A field product and a difference operator are naturally
extended as

\vspace{-6mm}
%
%%%%%%%%%%%%%%%%
\begin{eqnarray}
(\phi \cdot \psi)_n^c 
  \equiv \sum_{l,m}\sum^{N}_{a,b=1} 
             C_{lmn}^{abc}\,\phi_{l}^{a} \psi_{m}^{b}\,,\quad
%\label{multi-product}\\
(D\phi)_n^b \equiv \sum_{m}\sum^{N}_{a=1}\, 
                     D_{mn}^{ab}\,\phi_{m}^{a}\, ,
\label{multi-product&diff}
\end{eqnarray}
%%%%%%%%%%%%%%%%
%

\vspace{-2mm}
\noindent
where $a, b, c$ denote flavor indices.
A Leibniz rule in multi-flavor systems can be expressed as

\vspace{-3mm}
%
%%%%%%%%%%%%%%%%
\begin{equation}
\sum_{k}\sum^{N}_{d=1}C^{abd}_{lmk}\,D^{dc}_{kn} 
   = \sum_{k}\sum^{N}_{d=1} C^{dbc}_{kmn}\,
        D^{ad}_{lk} 
    + \sum_{k}\sum^{N}_{d=1} C^{adc}_{lkn}\,
        D^{bd}_{mk}\, .
\label{multi-leibniz}
\end{equation}
%%%%%%%%%%%%%%%%
%

\vspace{-1mm}
\noindent
The translation invariance and the locality for
$C^{abc}_{lmn}$ and $D^{ab}_{mn}$ are defined in the same
way as in single flavor systems, and lead to the
holomorphic functions

\vspace{-6mm}
%
%%%%%%%%%%%%%%%%
\begin{eqnarray}
\hat{C}^{abc}(v,w) \equiv
  \sum^{\infty}_{l,m=-\infty} C^{abc}(l,m)\, v^{l} w^{m},\quad
%\label{multi-hol-C}\\
\hat{D}^{ab}(z) \equiv
  \sum^{\infty}_{m=-\infty} D^{ab}(m)\, z^{m}.
  \label{multi-hol}
\end{eqnarray}
%%%%%%%%%%%%%%%%%
%

\vspace{-2mm}
One might hope that $N$-flavor systems could evade the
no-go theorem proved in one-flavor systems but the theorem
turns out to still hold.
This is because the proof can reduce to the one-flavor case
by diagonalizing $\hat{D}^{ab}(z)$ such that
$\hat{D}^{ab}(z)=\delta^{ab}\,\hat{D}^{a}(z)$.
Thus, a multi-flavor extension seems not to work to
solve our problem.
However, there exists a loophole to escape from the
no-go theorem.
A key observation is that a linear combination of an
{\it infinite} number of holomorphic functions is not
necessarily holomorphic.
The holomorphy of $\hat{C}^{abc}(v,w)$ and $\hat{D}^{ab}(z)$
is not necessarily preserved in diagonalizing $\hat{D}^{ab}(z)$
with infinite flavors, and hence the proof for one-flavor
cannot be applied to an infinite number of flavors.

In fact, we find a solution to the equation (\ref{multi-leibniz})
\cite{Taming}:

\vspace{-6mm}
%
%%%%%%%%%%%%%%%%
\begin{eqnarray}
C^{abc}_{lmn} = \delta_{l-n,b}\,\delta_{n-m,a}\,
                \delta_{a+b,c}\,,\quad
%  \label{multi-sol-product}\\
D^{ab}_{mn} = d(a-b)\,\big( \delta_{m-n,a-b}
               - \delta_{m-n,-(a-b)}\big)\,,
  \label{multi-sol}
\end{eqnarray}
%%%%%%%%%%%%%%%%%
%
where $d(a-b)$ is an arbitrary function of $a-b$.
The characteristic features of the above solution are
listed below:
They are translationally invariant because $D^{ab}_{mn}$
$(C^{abc}_{lmn})$ depends only on $m-n$ ($l-n$ and $m-n$).
They are local because the functions
%
%%%%%%%%%%%%%%%%
\begin{eqnarray}
\hat{C}^{abc}(v,w) =
  \delta_{a+b,c}\,v^{b}\,w^{-a}\,,\quad
%  \label{multi-sol-hol-product}\\
\hat{D}^{ab}(z) =
  d(a-b)\,\big( z^{a-b} - z^{-(a-b)}\big)
  \label{multi-sol-hol}
\end{eqnarray}
%%%%%%%%%%%%%%%%%
%
are holomorphic on the annulus domain.
The expressions (\ref{multi-sol}) imply nontrivial connections
between the lattice sites and the flavor indices and
the necessity of an infinite number of flavors.
We could diagonalize $\hat{D}^{ab}(z)$ given in 
eq.(\ref{multi-sol-hol}) but the resulting expression
turns out to become non-holomorphic.
Thus, we cannot apply the previous proof in the present
case, as mentioned before.
The solution (\ref{multi-sol}) is local in the space direction, 
as they should be, but they are, in some sense,
\lq\lq non-local\rq\rq\ in the flavor direction.
Thus, one might say that the non-locality problem can be
resolved by transferring the non-locality in the space
direction to the flavor direction.

The expressions (\ref{multi-sol}) may become more tractable by
writing them into a matrix form.
A lattice field $\phi^{a}_{n}$ will be replaced by
a matrix $\Phi_{ij}$ with the identification of
$a=i-j$ and $n=i+j$.
The difference operator $(D\phi)^{a}_{n}$ turns out
to be a commutator $[d,\Phi]_{ij}$, where the matrix
$d$ is given by $d_{ij} \equiv d(i-j)$.
The field product $(\phi\cdot\psi)^{a}_{n}$ is found
to be just the product of matrices,
$(\Phi\Psi)_{ij} = \sum_{k}\Phi_{ik}\Psi_{kj}$.
The Leibniz rule $D(\phi\cdot\psi) = (D\phi)\cdot\psi
+ \phi\cdot(D\psi)$ is rather trivial in the matrix
representation and is replaced simply by a trivial
commutator relation $[d,\Phi\Psi] = [d,\Phi]\Psi
+ \Phi [d,\Psi]$.
The summation over the lattice sites and the flavor
indices is replaced by the trace of matrices.
%Thus, the replacement by the matrix representation
%is summarized as follows:
%%
%%%%%%%%%%%%%%%%%
%\begin{eqnarray}
%%
%\phi^{a}_{n}\ \ &\longrightarrow&
%  \ \ \Phi_{ij}\quad (a-i-j,\ n=i+j)\,,
%  \label{matrix_representation1}\\
%(D\phi)^{a}_{n}\ \ &\longrightarrow&
%  \ \ [d, \Phi]_{ij}\,,
%  \label{matrix_representation2}\\
%(\phi\cdot\psi)^{a}_{n}\ \ &\longrightarrow&
%  \ \ (\Phi\Psi)_{ij} = \sum_{k}\Phi_{ik}\Psi_{kj}\,,
%  \label{matrix_representation3}\\
%D(\phi\cdot\psi)=(D\phi)\cdot\psi+\phi\cdot(D\psi)\ \ 
% &\longrightarrow&
%  \ \ [d,\Phi\Psi] = [d,\Phi]\Psi + \Phi [d,\Psi]\,,
%  \label{matrix_representation4}\\
%\sum_{n}\sum_{a}\ \ &\longrightarrow&
%  \ \ {\textrm{tr}} [\qquad]\,.
%  \label{matrix_representation5}
%%
%\end{eqnarray}
%%%%%%%%%%%%%%%%%%
%

%
%
%
%%%%%%%%%%%%%%%%%%%%%%%%%%%%%%%%%%%%%%%%%%%%%%%%%%%%%%%%%%%%%%%%
%%%%%%%%%%%%%%%%%%%%%%%%  Section 4  %%%%%%%%%%%%%%%%%%%%%%%%%%%
%%%%%%%%%%%%%%%%%%%%%%%%%%%%%%%%%%%%%%%%%%%%%%%%%%%%%%%%%%%%%%%%
\section{$N=2$ SUSY QM on the Lattice}
%%%%%%%%%%%%%%%%%%%%%%%%%%%%%%%%%%%%%%%%%%%%%%%%%%%%%%%%%%%%%%%%
%
%
%
Since we have succeeded to find a local difference operator
and a local field product equipped with the Leibniz rule,
we can construct supersymmetric lattice models.
In this section, as an example, we present a lattice model
of $N=2$ supersymmetric quantum mechanics, and discuss
various supersymmetric properties.
The details will be reported in a forthcoming paper.

Our lattice action of $N=2$ supersymmetric quantum
mechanics equipped with the tools developed in the 
previous section is\footnote{
The action was originally found in Ref.\cite{KSS05}.
}
%
%%%%%%%%%%%%%%%%
\begin{eqnarray}
S = {\textrm{tr}} \Big[\ 
       \frac{1}{2} [d,q]^{2} - \frac{1}{2} F^{2} + \bar{\psi} [d,\psi]
       + \lambda F q^{2} + \lambda \big( \bar{\psi} q \psi
        + \bar{\psi} \psi q \big)\,\Big]\,,
  \label{action}
\end{eqnarray}
%%%%%%%%%%%%%%%%%
%
where $d$ is an anti-hermitian matrix and $q, F$ are bosonic
matrices and $\psi, \bar{\psi}$ are fermionic ones.
We should be careful about the order of the fields because
they are represented by matrices.
It is easy to see that the lattice action is invariant under
the supersymmetry transformations
%%
%%%%%%%%%%%%%%%%%
%\begin{eqnarray}
%%
%\delta q &=& \varepsilon \bar{\psi} - \bar{\varepsilon}\psi\,,
% \nonumber \label{SUSY_trasf_q}\\
%\delta \psi &=& \varepsilon ( -[d,q] + F )\,,
% \nonumber  \label{SUSY_trasf_psi}\\
%\delta \bar{\psi} &=& \bar{\varepsilon} ( [d,q] + F )\,,
% \nonumber  \label{SUSY_trasf_psi*}\\
%\delta F &=& \varepsilon [ d, \bar{\psi}] 
%             + \bar{\varepsilon} [ d, \psi]\,,
%  \label{SUSY_trasf}
%%
%\end{eqnarray}
%%%%%%%%%%%%%%%%%%
%
%
%%%%%%%%%%%%%%%%
\begin{eqnarray}
\delta q = \varepsilon \bar{\psi} - \bar{\varepsilon}\psi,\ \ 
% \nonumber \label{SUSY_trasf_q}\\
\delta \psi = \varepsilon ( -[d,q] + F ),\ \ 
% \nonumber  \label{SUSY_trasf_psi}\\
\delta \bar{\psi} = \bar{\varepsilon} ( [d,q] + F ),\ \ 
% \nonumber  \label{SUSY_trasf_psi*}\\
\delta F = \varepsilon [ d, \bar{\psi}] 
             + \bar{\varepsilon} [ d, \psi],
  \label{SUSY_trasf}
\end{eqnarray}
%%%%%%%%%%%%%%%%%
%
where $\epsilon$ and $\bar{\varepsilon}$ are infinitesimal
grassmann parameters (but not matrices).
The invariance implies that the action possesses the full
exact $N=2$ supersymmetry.
This is the most important property of our model.
We can rewrite the lattice action (\ref{action}) in the
superfield formulation, such that
%
%%%%%%%%%%%%%%%%
\begin{eqnarray}
S = \int \!\! d\bar{\theta}d\theta\ {\textrm{tr}} \Big[\ 
       \frac{1}{2} (\bar{D}\Phi) (D\Phi) + \frac{\lambda}{3}
        \Phi^{3}\,\Big]\,,
  \label{superfield_action}
\end{eqnarray}
%%%%%%%%%%%%%%%%%
%
where $\Phi$ and $D, \bar{D}$ denote the superfield and the
supercovariant derivatives.
%defined by
%%
%%%%%%%%%%%%%%%%%
%\begin{eqnarray}
%%
%\Phi(\theta,\bar{\theta}) &=&
%   q + \theta\bar{\psi} - \bar{\theta} \psi 
%    + \theta\bar{\theta} F\,,
%  \label{superfield}\\
%D \Phi(\theta,\bar{\theta}) &=&
%   i \frac{\partial\Phi(\theta,\bar{\theta})}{\partial\bar{\theta}}
%    + i\theta\,[d,\Phi(\theta,\bar{\theta})] \,,
%  \label{supercovariant}\\
%\bar{D} \Phi(\theta,\bar{\theta}) &=&
%   i \frac{\partial\Phi(\theta,\bar{\theta})}{\partial\theta}
%    + i\bar{\theta}\, [d,\Phi(\theta,\bar{\theta})] \,.
%  \label{supercovariant*}
%%
%\end{eqnarray}
%%%%%%%%%%%%%%%%%%
%
Furthermore, the action can be written into a $Q$-exact form
\cite{Witten88}
%
%%%%%%%%%%%%%%%%
\begin{eqnarray}
S = \bar{Q} Q\ {\textrm{tr}} \Big[\ 
       \frac{1}{2} \bar{\psi}\psi + \frac{\lambda}{3}q^{3}
        \,\Big]
  = -  Q\bar{Q}\ {\textrm{tr}} \Big[\ 
       \frac{1}{2} \bar{\psi}\psi + \frac{\lambda}{3}q^{3}
        \,\Big]\,,
  \label{Q_exact}
\end{eqnarray}
%%%%%%%%%%%%%%%%%
%
where $Q$ and $\bar{Q}$ denote the supercharges, which generate
the supersymmetry transformations (\ref{SUSY_trasf}).
The other interesting feature is that the action has two kinds
of the Nicolai mappings \cite{Nicolai_map, CG83, Sakai-Sakamoto83} 
given by

\vspace{-8mm}
%
%%%%%%%%%%%%%%%%
\begin{eqnarray}
\xi = [d,q] + \lambda q^{2}\,,\quad
%  \label{Nicolai_map1}\\
\eta = -[d,q] + \lambda q^{2}\,.
  \label{Nicolai_map}
\end{eqnarray}
%%%%%%%%%%%%%%%%%
%
In terms of the two Nicolai mappings, the on-shell action can be
expressed in two ways:
%
%%%%%%%%%%%%%%%%
\begin{eqnarray}
S^{on-shell}
 = \frac{1}{2}\xi_{ij}\xi_{ji} 
    + \bar{\psi}_{ij}\frac{\partial\xi_{ji}}{\partial q_{kl}}\psi_{kl}
 = \frac{1}{2}\eta_{ij}\eta_{ji} 
    + \bar{\psi}_{ij}\frac{\partial\eta_{lk}}{\partial q_{ij}}\psi_{kl}\,.
  \label{onshell_action}
\end{eqnarray}
%%%%%%%%%%%%%%%%%
%
It follows from the above expressions that each of the two
Nicolai mappings $\xi$ and $\eta$ are found to be connected 
with one of the two supercharges of the $N=2$ supersymmetry
\cite{PS82,CK02}.

We should make a comment on the doubling problem.
The lattice action (\ref{action}) unfortunately suffers 
from the doubling in both the spectrum of bosonic and fermionic
variables.
We can, however, add a supersymmetric Wilson term to the 
lattice action (\ref{action}) to avoid the doubling problem.

It should be noting that the lattice action is exactly
supersymmetric even for a finite size of the matrices.
This does not, however, contradict the no-go theorem 
because the theorem has been proved only for an infinite
lattice volume.
In fact, the locality is obscure for a finite lattice.

%
%
%
%%%%%%%%%%%%%%%%%%%%%%%%%%%%%%%%%%%%%%%%%%%%%%%%%%%%%%%%%%%%%%%%
%%%%%%%%%%%%%%%%%%%%%%%%  Section 5  %%%%%%%%%%%%%%%%%%%%%%%%%%%
%%%%%%%%%%%%%%%%%%%%%%%%%%%%%%%%%%%%%%%%%%%%%%%%%%%%%%%%%%%%%%%%
\section{Summary and Discussions}
%%%%%%%%%%%%%%%%%%%%%%%%%%%%%%%%%%%%%%%%%%%%%%%%%%%%%%%%%%%%%%%%
%
%
%
We have proved the no-go theorem that it is impossible to 
construct a lattice field theory in an infinite
lattice volume with any field products
and difference operators that satisfy the
following three properties:
(i) translation invariance, (ii) locality and (iii) Leibniz rule.
We then proposed a way to escape from the no-go theorem by
introducing infinite flavors and presented a matrix realization
of the lattice formulation equipped with the Leibniz rule.
In terms of the tools in the matrix realization, we constructed
a lattice model of $N=2$ supersymmetric quantum mechanics
and discussed various interesting properties.
One of the remarkable features is that the model realizes
the full exact supersymmetry but most of known
supersymmetric lattice models can, in contrast, possess 
supersymmetry at most partially.\footnote{
Other lattice models that realize exact supersymmetry
have been proposed in Ref.\cite{Hokudai}.
However, a breakdown of the supersymmetry at the quantum
level was reported \cite{Nagata08}.
Another interesting idea to construct lattice models with an exact
fermionic symmetry %on an Ichimatsu lattice 
has been presented 
in Ref.\cite{ichimatsu}, though the relation between the 
fermionic symmetry and supersymmetry is unclear.
}

Since our lattice models include an infinite number of flavors,
we need to reduce the number of flavors in the continuum limit
at least at low energies.
%One idea is to keep only finite flavors and discard the
%others by hand.
%We could then have lattice models of finite degrees of freedom
%with partial supersymmetry.
An attractive idea is to use an analogy between infinite
flavors of our models and Kaluza-Klein modes on extra dimensions.
If we can add \lq\lq KK mass\rq\rq\ terms in our lattice actions
in a supersymmetric way, only finite degrees of freedom could
survive at low energies.

In the previous section, we have presented a $0+1$-dimensional
supersymmetric lattice model.
The extension to higher dimensions will be straightforward
for Wess-Zumino type models.
In fact, we have succeeded to construct lattice models
of $d=2$ $N=2$ and $d=4$ $N=1$ Wess-Zumino models in our
formulation.
The introduction of gauge fields is, however, less trivial.
Our lattice models have a genuine non-commutative nature
and it seems to be natural to embed vector indices 
(as well as spinor indices) in matrices.

An advantage of our lattice formulation is that the lattice
models are exactly supersymmetric even for a finite lattice
size, and hence numerical simulations can be done.

The full details of the supersymmetric lattice models given
in Section 4 and further progress will be reported
in a forthcoming paper.

%
%
%
%%%%%%%%%%%%%%%%%%%%%%%%%%%%%%%%%%%%%%%%%%%%%%%%%%%%%%%%%%%%%%%%
%%%%%%%%%%%%%%%%%%%%%%%%  references  %%%%%%%%%%%%%%%%%%%%%%%%%%
%%%%%%%%%%%%%%%%%%%%%%%%%%%%%%%%%%%%%%%%%%%%%%%%%%%%%%%%%%%%%%%%

%%%%%%%%%%%%%%%%%%%%%%%%%%%%%%%%%%%%%%%%%%%%%%%%%%%%%%%%%%%%%%%%
%
%
%
%
%
%%%%%%%%%%%%%%%%%%%%%%%%%%%%%%%%%%%%%%%%%%%%%%%%%%%%%%%%%%%%%%%%
%%%%%%%%%%%%%%%%%%%%%%%%%%%%%%%%%%%%%%%%%%%%%%%%%%%%%%%%%%%%%%%%
\end{document}